\documentclass[11pt,a4paper]{scrartcl}

\usepackage{CLICdp}

\usepackage{CLICdp_definitions}

\usepackage{tabulary}
\newlength{\abc}
\AtBeginDocument{%
\renewcommand{\ref}[1]{\mbox{\autoref{#1}}}

}

\title{From precision physics to the energy frontier\\ with the Compact Linear Collider}

\date{\today}

\addauthor{Eva Sicking}{\institute{1}\hcomma\editor{Corresponding author: \href{mailto:eva.sicking@cern.ch}{eva.sicking@cern.ch}}}
\addauthor{Rickard Str\"{o}m}{\institute{1}}

\addinstitute{1}{CERN, Switzerland}

\abstract{The Compact Linear Collider (CLIC) is a proposed high-luminosity collider that would collide electrons with their antiparticles, positrons, at energies ranging from a few hundred Giga-electronvolts (GeV) to a few Tera-electronvolts (TeV). By covering a large energy range and by ultimately reaching multi-TeV $\Pep\Pem$ collisions, scientists at CLIC aim to improve the understanding of nature's fundamental building blocks and to discover new particles or other physics phenomena. CLIC is an international project with institutes world-wide participating in the accelerator, detector and physics studies. First $\Pep\Pem$ collisions at CLIC are expected around 2035, following the High-Luminosity phase of the Large Hadron Collider at CERN.}

\titlecomment{The article presents an up-to-date overview of the CLIC project for a general audience\\ as presented to the 2020 update of the European Strategy for Particle Physics.} 

\graphicspath{ {./logos/}{./figures/} }

\begin{document}

% generates the title page
\titlepage

% include source for sections
\section{CLIC as the next high-energy collider }
The discovery of the Higgs boson at the Large Hadron Collider (LHC) in 2012 [1,2] was an important milestone in high-energy physics. 
It completes a puzzle that scientists have been working on for decades: 
the Standard Model (SM) of Particle Physics. However it is known that there must be physics Beyond the Standard Model (BSM), for example to account for dark matter and the matter-antimatter asymmetry in the Universe. 
In the absence of clear guidance towards the scale of BSM physics from the LHC and other experiments to date, plans are already underway to prepare for the next-generation project in high-energy physics.

The Compact Linear Collider (CLIC) is a proposed high-luminosity $\Pep\Pem$ collider [3--6]. 
It is currently the only mature lepton-collider project aiming for multi-TeV energies. 
CLIC is an international project hosted by CERN with 75 collaborating institutes involved in the accelerator, detector and physics studies. 
The advantage of CLIC as the next collider is that it gives access to two complementary search paths for new physics. 
The first path focuses on studying known SM processes with unprecedented precision, to search for deviations from the predicted behaviour. 
Such deviations would represent indirect evidence of BSM physics. 
The second path searches for direct production of new particles. 
The clean environment of $\Pep\Pem$ collisions is favourable, both for the detection of even tiny deviations of the expected SM properties and for the detection of rare new signals [5]. 
Combined, these results from CLIC would provide important guidance for particle physics.

The LHC accelerates and collides protons, composite particles made of quarks and gluons, which each carries a varying fraction of the proton's energy. 
The complex structure of the proton limits knowledge of the individual quarks or gluons de facto participating in the collisions, complicating interpretation of the data. 
In addition, proton-proton collisions produce vast rates of backgrounds induced by strong interactions between quarks and gluons. 
Interesting interactions need to be filtered from these background events using triggers. 
Electrons and positrons, on the other hand, are elementary particles: the colliding system is well defined in terms of particle type, energy and polarisation, without large backgrounds. 
In addition, the generally lower cross sections in $\Pep\Pem$ collisions combined with instantaneous luminosities similar to those at the LHC lead to a lower event rate. 
These characteristics and the low duty cycle of linear colliders make a trigger-less detector readout possible at CLIC. 
Radiation levels are significantly lower at $\Pep\Pem$ colliders, relaxing constraints in terms of radiation hardness in detector design. 
Linear colliders such as CLIC provide high levels of electron-beam polarisation at all energies, which for example might help to characterise newly discovered particles or processes in detail.

Reaching TeV-scale energies in an $\Pep\Pem$ collider is a challenging endeavour. 
In ring-shaped particle accelerators, such as the LHC, the circulating particles of mass m lose a fraction of their energy E via synchrotron radiation when being forced to follow a circular trajectory. 
The resulting energy loss, following $\Delta E\sim (E/m)^4$, increases with higher energies and smaller particle masses. 
Electrons, almost 2000 times lighter than protons, are therefore especially prone to synchrotron radiation. 
Linear colliders, where synchrotron radiation is naturally avoided, provide an efficient solution for reaching high $\Pep\Pem$ collision energies.

In circular accelerators, the ultimate collision energy is reached by circulating the beam through few acceleration units, at each turn increasing the particle energy by a small amount. 
In a linear accelerator, however, the full collision energy must be delivered to the particles in one passage through the accelerator. 
In this case the accelerator needs to be equipped with many acceleration structures distributed along its length. 
In order to keep the scale of a high-energy linear collider project within reasonable limits in terms of length and cost, very high acceleration gradients are required. 
\section{An innovative acceleration technology to reach multi-TeV energies}
\begin{figure}
 \centering
 \includegraphics[width=\textwidth]{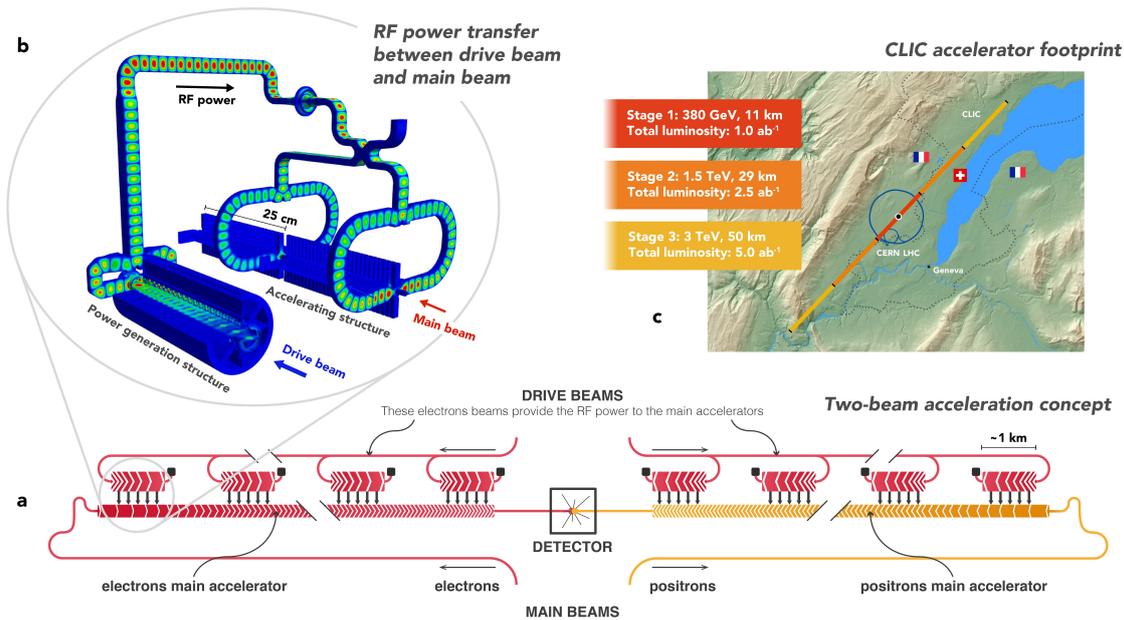}
\caption{
{\bfseries The CLIC acceleration scheme and accelerator footprint.} 
{\bfseries a}, Particle acceleration at CLIC is achieved with a two-beam acceleration concept, in which a high-current (100\,A), low-energy drive beam runs in parallel to a low-current (1.2\,A), high-energy main beam of colliding particles (image credit: CERN). 
{\bfseries b}, the kinetic energy of the drive beam, which is decelerated from 2.4\,GeV to 240\,MeV, is converted into radio-frequency (12\,GHz) power that is used to accelerate the main beam, from 9\,GeV up to 1.5\,TeV [7]. 
{\bfseries c}, the footprint of the CLIC main linear accelerator is illustrated for a possible implementation at the research centre CERN in Geneva, Switzerland (image credit: CERN). }
\label{fig:acc}
\end{figure}

CLIC is proposed to be built and operated in three stages with increasing collision energy and luminosity, providing collisions in a wide range of centre-of-mass energies; from 350--380\,GeV up to 3\,TeV [4]. 
\ref{fig:acc} shows the CLIC acceleration concept, a two-beam acceleration unit, as well as the footprint of the different CLIC stages for a possible implementation at the research centre CERN in Geneva, Switzerland.

The ability of CLIC to reach TeV energies in a compact and cost-effective way is one of its most important design features and unique among the proposed concepts for future electron-positron colliders. 
It is made possible by a novel two-beam acceleration technology, illustrated in \ref{fig:acc}, which enables efficient acceleration of the colliding beams using normal conducting accelerating structures that achieve an acceleration gradient of $\sim$100 million volts per metre (MV/m) [4]. 
The radio-frequency pulses needed to accelerate the main beams are generated by decelerating a second, high-intensity electron ``drive'' beam in dedicated power-extraction structures. 
An animation of the CLIC accelerator concept can be found at~[8].

Detailed simulation studies have concluded that the optimal collision energy for the first stage of CLIC is 380\,GeV [9], giving access to both precision Higgs-boson and top-quark physics, with a targeted integrated luminosity of $1\,\text{ab}^{-1}$ [3].
Neither of these particles has previously been studied in $\Pep\Pem$ collisions. 
The two-beam accelerating scheme allows for a well-defined upgrade path towards multi-TeV collision energies [4]; 
the two higher energy stages, at 1.5\,TeV and 3\,TeV, would accumulate $2.5\,\text{ab}^{-1}$ and $5\,\text{ab}^{-1}$ of integrated luminosity, respectively [3]; 
see \ref{tab:box} for further details. 
A clear advantage of the staged programme of CLIC is its flexibility: 
the collision energy or schedule of operation can be adapted in light of any indication of new physics. 
Note further that, in contrast to circular colliders, the luminosity at a linear collider increases with centre-of-mass energy, making high-energy operation at CLIC particularly advantageous.

First collisions at CLIC could be realised as early as 2035 [3], allowing continued exploration of the electroweak energy scale at the end of the High-Luminosity phase of the LHC programme (HL-LHC) [10]. 
With 7--8 years of operation per stage and an additional 2 years of construction and commissioning between stages, CLIC is expected to operate for about 30 years. 

Technical development, optimisation and system tests of the accelerator design have resulted in significant progress in recent years. 
Major investments have been made in the high-power 12\,GHz radio-frequency (X-band) linac technology, the main pillar of the two-beam acceleration scheme. 
Long-term operation of accelerating gradients exceeding 100\,MV/m has been routinely demonstrated in test stands at CERN, KEK and SLAC [4]. 
Further, the ambitious luminosity goals of CLIC put stringent requirements on alignment and stability as well as on the quality of the nanometre-sized beams. 
The required beam quality has been successfully demonstrated in several modern synchrotron light facilities [4]. 
The growing use of CLIC-like technology in accelerator projects (e.g.\ light sources) worldwide is particularly important when moving towards industrialisation of the production of CLIC accelerator components. 
A detailed discussion of the CLIC technology choices and challenges can be found in [4,11].

\begin{table}[h!]
\caption{The CLIC baseline performance and special operating conditions} 
\label{tab:box}
\footnotesize
\fbox{
\begin{tabulary}{\linewidth}{L|L}
{\bfseries Beam structure}\newline
In the CLIC main beam, particles are grouped together in bunches with several billion electrons or positrons per bunch, with a spacing of 0.5\,ns from one bunch to the next. 
So called "bunch trains" are composed of several hundred bunches that arrive at the detector with a baseline repetition rate of 50\,Hz [4,6]. 
The beam size at the interaction point will be 149\,nm (40\,nm) times 2.9\,nm (1\,nm) in the transverse direction of the beam and 70\,$\upmu$m (44 \,$\upmu$m) along the beam direction for operation at 380\,GeV (3\,TeV) [4].       
& 
{\bfseries Luminosity and beam energy}\newline
The instantaneous luminosity of CLIC increases as a function of the centre-of-mass energy with values of 1.5, 3.7 and $5.9 \times 10^{34}\,\text{cm}^{-2}\text{s}^{-1}$ for the three proposed energy stages, 380\,GeV, 1.5\,TeV and 3\,TeV [3]. 
The final beam energy spread at CLIC amounts to 0.35\% at all collision energies [3]. 
The beam energy can be measured with high precision using processes such as $\Pep\Pem\to\PGmp\PGmm\PGg$, for example to less than 10\,MeV at 350\,GeV.
\\
\midrule
{\bfseries Luminosity upgrade}\newline
For the initial energy stage, CLIC operation at 100 Hz bunch train repetition rate is under discussion, which is a doubling of the baseline collision rate. Furthermore, multiple strategies for reducing the vertical beam emittance, which have the potential to substantially increase the luminosity further, are under study [13]. 
& 
{\bfseries Detector design in view of the beam structure}\newline
The 20 ms gap between bunch trains allows switching off much of the detector electronics between trains. 
This ``power pulsing'' reduces the required cooling, thereby enabling lighter or more compact sub-detectors [6]. 
Energy deposits from beam-induced background events produced within a fraction of the bunch train could make it difficult to reconstruct an interesting collision event. 
Such effects are mitigated efficiently by dedicated reconstruction algorithms, making use of high spatial granularities and nanosecond time resolution in the detector [6].\\
\midrule
{\bfseries Key detector performance parameters}\newline
The following physics-driven detector requirements are generally realised by full simulations of the CLIC detector concept [3,12]:\newline
- Track-momentum resolution for high momentum tracks in the barrel detector of $\sigma_{p_T}/p_T^2 \leq 2\times 10^{-5}\,\text{GeV}^{-1}$;\newline
- Impact parameter resolution of $\sigma_{d_0}^2 = (5 \upmu$m$)^2 + (15 \upmu$m$\,\text{GeV})^2/ (p^2 / \sin^3 \theta)$;\newline
- Jet-energy resolution for light quark jets of $\sigma_E/E \leq 3.5\%$ for jet energies in the range 100\,GeV to 1\,TeV and 5\% at 50\,GeV.
& 
{\bfseries Interaction points and experiments}\newline
The current CLIC baseline features a single interaction point with one experiment to which the full luminosity is delivered.
An alternative option is to operate two detectors in turns, in a so-called push-pull mode [14]. 
A second alternative would be to construct two beam-delivery systems and two interaction points sharing the full luminosity [13].
\\
\midrule
{\bfseries Advantage of operating at high energy}\newline
Operation of the CLIC accelerator at centre-of-mass energies not included in the project baseline staging scenario, such as at 91\,GeV (Z threshold) [13] or at 250\,GeV ($\PZ\PH$ threshold), is technically possible if motivated by physics.
It is important to note that in many BSM scenarios the contributions to processes like $\Pep\Pem \to \PW\PW$, $\Pep\Pem \to \PZ\PH$ and $\Pep\Pem \to \PQt\PAQt$ rise strongly with the centre-of-mass energy. 
High-energy CLIC operation can therefore provide better sensitivity to new physics from these processes compared to operation at their respective production thresholds despite the smaller event numbers [5].
Furthermore CLIC's high-energy operation gives access to large Higgs-boson samples from $\PW\PW$-fusion events, double Higgs-boson production as well as enables direct searches up to the kinematic limit of the collider. 
& 
\end{tabulary} 
}
\end{table}

\section{A multi-purpose detector for precision physics}

\begin{figure}[h!]
 \centering
 \includegraphics[width=\textwidth]{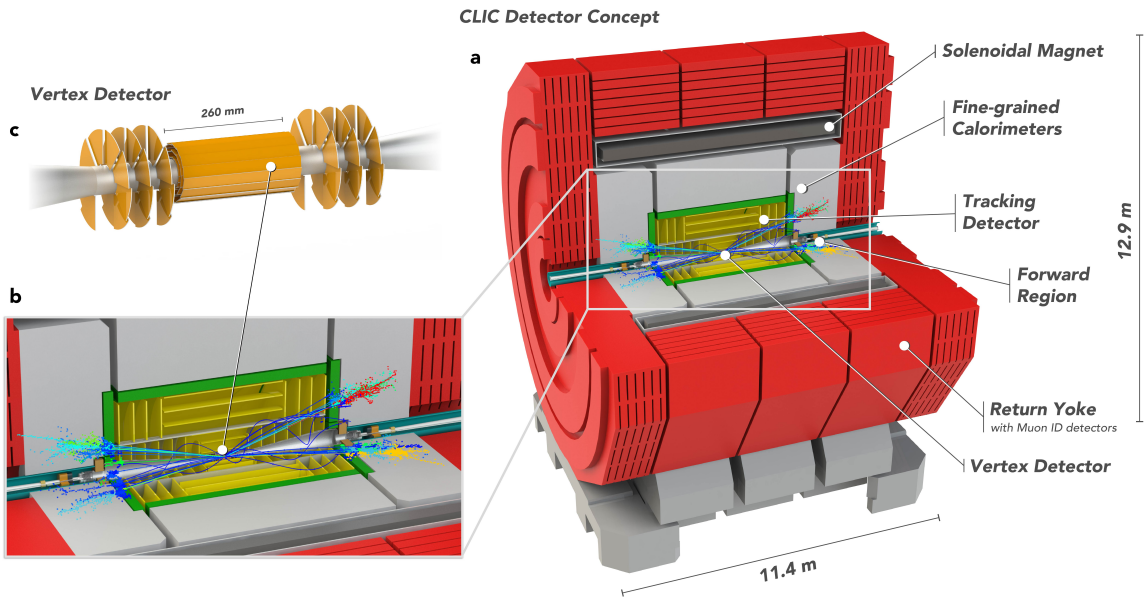}
\caption{
{\bfseries The CLIC detector concept with inlays showing the inner detector regions in greater detail.}
{\bfseries a}, the detector has a diameter of 12.9\,m and a side length of 11.4\,m. 
The sub-detector layers are arranged hermetically around the collision point in a cylindrical configuration with two endcaps. 
Charged particle tracks and energy deposits from a simulated double Higgs boson event ($\Pep\Pem \to\PH\PH \PGne\PAGne \to \PQb\PAQb\PQb\PAQb \PGne\PAGne$) at a centre-of-mass energy of 3\,TeV are indicated. 
These are shown after subtraction of energy deposits from out-of-time backgrounds generated in other collision events within the same 156\,ns long bunch train (see \ref{tab:box} for more details). 
The background subtraction is made possible due to the high spatial granularity of the detector as well as due to its hit time resolution reaching as low as a few nanoseconds per readout unit. 
The four bottom-quarks hadronise resulting in four jets. 
Additional views of the tracking, {\bfseries b}, and vertex, {\bfseries c}, detectors are also shown (image credit: CERN).}
\label{fig:det}
\end{figure}

CLIC features an advanced detector concept for high-precision measurements and searches [6,12], shown in \ref{fig:det}. 
Modern detector technologies, profiting in particular from the rapid advancement in the silicon industry, as well as detector designs adapted to the favourable experimental conditions at CLIC allow for a highly granular detector system with superior measurement precision, compared to the detectors implemented in the current LHC experiments. 

The two innermost sub-detectors of the CLIC detector, both based on silicon-pixel technology, are essential for the measurement of charged particles: 
a light-weight vertex detector with very small pixels, optimised for the reconstruction of the particle trajectories and their origins (vertices), is surrounded by a large tracking detector, essential for accurate reconstruction of the charged particle momenta. 
The CLIC detector further comprises sampling calorimeters for electromagnetic and hadronic showers, ECAL and HCAL, for the measurement of the particle energy; active detection layers of high granularity are interleaved with dense absorber layers, either tungsten or steel. 
The active calorimeter layers are equipped with silicon sensors for the ECAL and scintillator tiles coupled to silicon-photomultipliers for the HCAL. 
These sub-detectors are located inside a superconducting solenoid magnet providing a magnetic field of 4\,T, itself surrounded by a detector for muon identification, embedded in the magnet return yoke. 
The muon detectors are based on resistive plate chambers. 
In total, the CLIC detector comprises almost 19 billion readout channels. 
It is optimised for so-called particle flow analysis (PFA) [15] that improves the energy reconstruction of particle jets (narrow particle sprays originating from quarks or gluons), by combining measurements from all sub-detectors. 

The CLIC detector concept was designed in view of physics performance and the experimental conditions at CLIC [6,12]; 
detailed detector simulations are being carried out covering a broad set of physics observables and using the full centre-of-mass energy range of CLIC [5]. 
These simulations are based on realistic modelling of the detector response and include overlay of beam-induced backgrounds at the level expected at CLIC. 
Such detector simulations are vital for a reliable estimation of the detector performance. 
The sub-detectors are adapted to the CLIC beam structure as described in \ref{tab:box}.

\begin{figure}[h!]
 \centering
 \includegraphics[width=\textwidth]{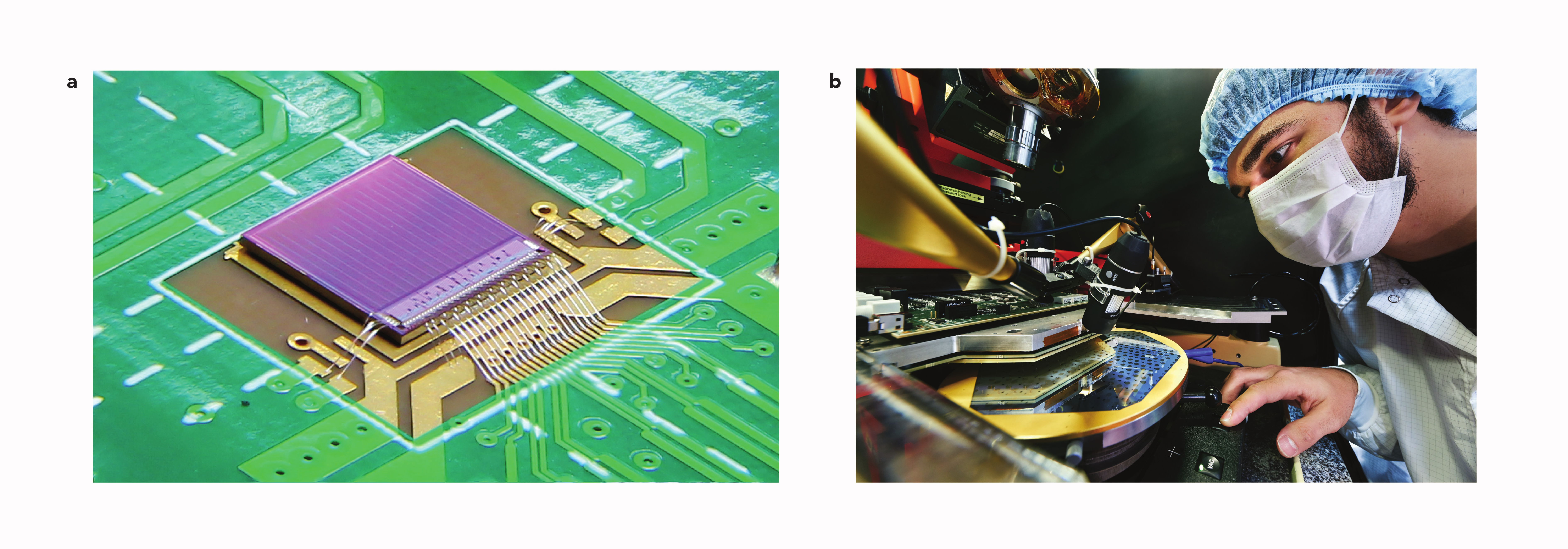}
\caption{
{\bfseries Silicon-based prototypes for tracking detectors and calorimeters.}
{\bfseries a}, a CLICTD silicon-pixel detector test chip, designed for the requirements of the CLIC tracking detector. 
The chip has a footprint of $5 \times 5\,\text{mm}^2$ and is implemented in monolithic CMOS technology, comprising both the sensor and the readout electronics. 
It contains a sensitive area segmented in 2048 readout channels. 
Each readout channel has a dimension of $30\times300\,\upmu$m$^2$ and is divided into 8 sub-pixels of $37.5 \times 3\,\upmu\text{m}^2$ size. 
The CLIC tracking detector will contain roughly $140\,\text{m}^2$ of silicon pixel detectors. 
The sub-pixel segmentation scheme tested with the CLICTD chip will help to reduce the number of readout channels for this large instrumented area, while maintaining a high measurement precision (image credit: CERN). 
{\bfseries b}, one of the 31,000 hexagonal silicon wafers foreseen for the CMS high-granularity calorimeter endcap upgrade. 
It is here being tested in a system for automated electrical characterisation before module assembly by a team of CMS and CLIC members (image credit: CERN). 
The sensors are made of 200\,mm diameter wafers with hundreds of individual pads of mostly 0.5--$1\,\text{cm}^2$ size [16]. 
This highly-granular calorimeter concept, initially studied for future linear collider detectors, is optimised for particle flow analysis to achieve an excellent jet-energy resolution [6,15]. 
The current CLIC detector design foresees an electromagnetic calorimeter with $2500\,\text{m}^2$ of silicon sensors and 100 million individual readout pads, which is 4 times the silicon surface and 16 times the cell number of the CMS endcap calorimeter upgrade [6].}
\label{fig:hardware}
\end{figure}

Technology demonstrators are built and validated for the most challenging sub-detectors of the CLIC detector, such as the vertex and tracking detectors and calorimeters [6]. 
To validate the performance, for instance in terms of energy or position resolution and detection efficiency, prototypes are tested with particle beams. 
\ref{fig:hardware}a shows a silicon-pixel detector test chip designed in monolithic CMOS technology for the CLIC tracking detector.

The CLIC detector development is performed in collaboration with other projects studying future collider detector concepts as well as in dedicated detector R\&D collaborations such as CALICE [17] and FCAL [18]. 
Most recently one of the proposed linear collider calorimeter concepts was adapted for the HL-LHC; 
the calorimeter endcap upgrade of the CMS detector will employ highly granular sampling calorimeters [16]. 
A silicon sensor developed for the CMS upgrade is shown in \ref{fig:hardware}b. 
Synergy exists between this detector concept and the calorimeters of the CLIC detector, for instance in view of detector calibration, integration and full system aspects.

\section{The CLIC physics programme}

\begin{figure}[h!]
 \centering
 \includegraphics[width=\textwidth]{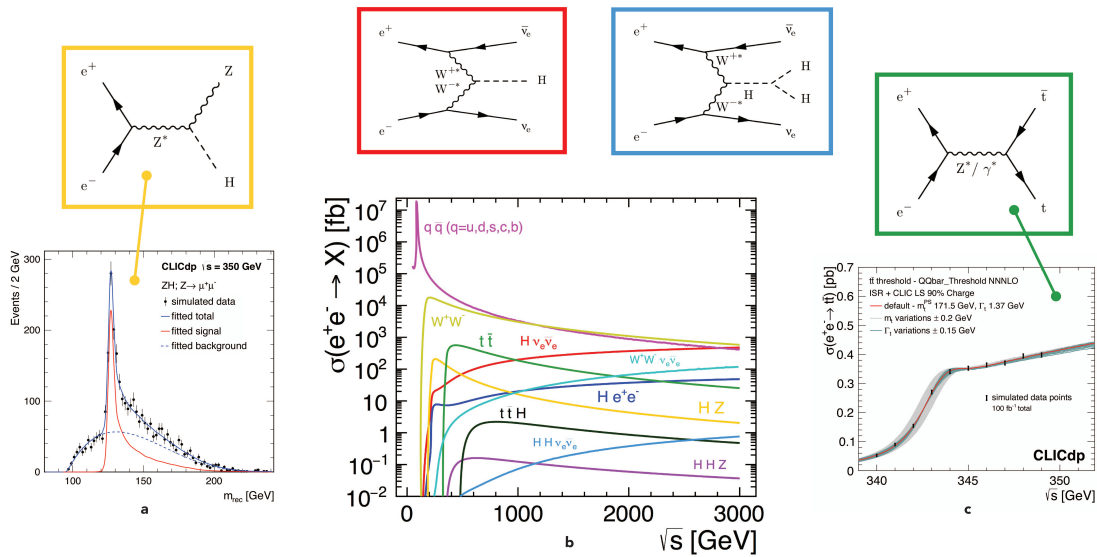}
\caption{
{\bfseries Highlighted Standard Model processes in the energy range of CLIC.}
{\bfseries b}, SM cross sections for key $\Pep\Pem$ processes (image credit: CLIC). 
Several particularly interesting channels are highlighted by showing their leading-order Feynman diagrams. 
Two example physics studies at the first energy stage are shown: 
{\bfseries a}, using Higgsstrahlung events ($\Pep\Pem\to\PZ\PH$) the Higgs boson can be accessed indirectly using the properties of the Z boson recoiling against it [19]; 
{\bfseries c}, the top-quark mass and other properties can be measured in a threshold scan at centre-of-mass energies near 350\,GeV [20].}
\label{fig:physics}
\end{figure}

The CLIC programme offers two complementary paths towards new physics. 
One path provides indirect access through precision measurements of already known processes and particles, such as the Higgs boson. 
The other path focuses on direct searches for new physics phenomena, such as the discovery of new particles. 
The first stage of CLIC, with a centre-of-mass energy of 380\,GeV, offers a precision physics programme focussing on measurements of the Higgs boson and the top quark [5]. 
Choosing 380\,GeV as the first centre-of-mass energy gives simultaneous access to several Higgs-boson production processes as well as top-quark pair production [9]. 
\ref{fig:physics} shows the SM cross sections for a number of additional processes at CLIC as a function of the collision energy.

The two main Higgs boson production channels at the first stage of CLIC are Higgsstrahlung, in which the Higgs boson is radiated from an off-shell Z boson ($\Pep\Pem\to\PZ\PH$), and Higgs-boson production via WW fusion ($\Pep\Pem\to\PH\PGne\PAGne$). 
The former is particularly interesting, since Higgs-boson candidate events can be identified in a model-independent way, solely via the properties of the Z boson recoiling against the Higgs boson, as shown in \ref{fig:physics}; 
no assumptions on the Higgs-boson decay are necessary. 
The recoil mass is estimated as $m^2_\text{rec} = s + m_{\PZ}^2 - 2 E_{\PZ} \sqrt{s}$ [19], where $\sqrt{s}$ is the centre-of-mass energy. 
Such a measurement is only possible at lepton colliders like CLIC and for example allows a model-independent upper limit well below 1\% to be set on the Higgs-boson decay to invisible particles.

Already at the initial stage of CLIC, the Higgs-boson width and many Higgs-boson couplings including the Higgs-boson coupling to charm quarks -- very difficult to observe at the LHC due to the large backgrounds -- could be measured with percent-level precision [19].
The study of charm-quark jets at CLIC is facilitated by the excellent flavour tagging capability of the CLIC detector as well as the low background levels. 
For specific Higgs-boson couplings such as the ones to bottom quarks, $\PZ$ bosons and $\PW$ bosons, CLIC would improve the precision significantly with respect to HL-LHC. 
For instance, these Higgs-boson coupling accuracies would be improved by a factor of $\sim$5 at the first CLIC stage, and by more than one order of magnitude in the full CLIC programme [21].

Furthermore, the first CLIC stage allows for detailed measurement of top-quark pair production including a dedicated threshold scan at centre-of-mass energies around the onset for top-quark pair production, at about 350\,GeV [20], as shown in \ref{fig:physics}. 
The top-quark mass is expected to be determined at CLIC with a statistical uncertainty of 20\,MeV; 
the total uncertainty is 50\,MeV and is dominated by current theory uncertainties [20]. 
The planned top-quark measurements further enable detailed studies of its electroweak couplings that are sensitive probes in the search for signs of new physics [5].

Profiting from the increase of both the $\PW\PW$-fusion cross section and the luminosity performance for increasing energies at linear colliders, the higher energy stages of CLIC, at 1.5\,TeV and 3\,TeV, would further improve the precision on the results discussed above and allow for detailed studies of rare Higgs-boson decays and additional Higgs-boson production channels [19]. 
The latter includes event topologies with multiple jets which can be measured with very high accuracy with the CLIC detector, such as the associated production of top quarks and a Higgs boson ($\Pep\Pem \to \PQt\PAQt\PH$), important for the accurate measurement of the top-Yukawa coupling. 
This also includes double-Higgs boson signatures, accessible mainly through $\PW$-boson fusion ($\Pep\Pem \to\PH\PH \PGne\PAGne$) and the double Higgsstrahlung process ($\Pep\Pem \to \PZ\PH\PH$), from which the Higgs boson trilinear self-coupling could be extracted at the 10\% level [22]. 
In comparison, the HL-LHC will reach an accuracy of only 50\% [23]. 
In both cases the Standard Model Higgs self-coupling is assumed. 
The self-coupling determines the shape of the Higgs potential, providing important insight into the mechanism of electroweak symmetry breaking. 
The precision reached at CLIC is at the level of the typical deviations induced by many new physics models [24].

The global impact of CLIC's precision measurements programme on its BSM reach is assessed within the so-called Effective Field Theory (EFT) framework. 
It efficiently allows knowledge from many different measurements at different energies to be combined and to be interpreted in terms of general new interactions added to the SM. 
It also recognizes the added value and complementarity between lepton colliders and hadron colliders. 
Combining the CLIC input from electroweak (for instance W- and fermion-pair production), Higgs-boson and top-quark processes, measured at the three energy stages, yields a significant improvement in the precision of EFT parameters and a substantial gain in the energy reach for indirect measurements, going well beyond the physics reach of HL-LHC [25]. 
The staged programme of CLIC, covering one order of magnitude in collision energy, is particularly advantageous for those EFT operators whose contributions grow with energy. 
Detailed studies show that the physics programme of CLIC would extend the reach for certain new physics effects to scales beyond 100\,TeV [25]. 
This allows to probe whether, for example, the Higgs boson has a substructure, and allows its dimension to be measured down to sizes of about 10--20\,m, corresponding to an energy scale of almost 20\,TeV [25]. 
The size is more than 100,000 times smaller than the proton radius and smaller by a factor of approximately 4 than the HL-LHC reach [25].

Direct discovery is often possible up to the kinematic limit of CLIC, for example half the centre-of-mass energy for pair-produced particles with electroweak-sized coupling strengths. 
CLIC's discovery potential was benchmarked with signatures from individual models of wide interest, such as models with additional spin-0 particles besides the SM Higgs boson or signatures with flavour changing neutral currents [5,14]. 
For many of these models, CLIC provides competitive sensitivity in a shorter time frame than other collider proposals [25].

For example, in Super-Symmetric (SUSY) extensions of the Standard Model, CLIC is especially competitive for electroweak states such as charginos, neutralinos and sleptons. 
This includes new particles with masses that are very close to each other in so-called compressed mass spectra which are difficult to observe at hadron colliders [5,25]. 

A wide range of dark matter models can be tested at CLIC, for instance through events with single photons and missing energy, a complementary approach to mono-jet searches at hadron colliders [5,25]. 
Searches for long-lived particles giving rise to disappearing tracks in the detector, will also benefit from the clean environment at CLIC and the excellent tracking detectors. 
A particularly interesting example of a new particle that could be discovered at CLIC are long-lived Higgsinos with a mass of 1.1\,TeV as predicted by certain models in which the Higgsino represents thermal dark matter [5,25].

If new particles are found at the HL-LHC or CLIC, their properties (mass, coupling, quantum numbers) can be accurately measured at CLIC, also aided by the ability to polarise the electron beam [5]. 
Dedicated threshold scans can be integrated into the CLIC running scenario also for these newly discovered particles.
\section{Synergies with far-future accelerator technologies}

The infrastructure for the next generation of linear colliders such as CLIC is also of interest for collider projects envisioned in the far future, for which new technologies such as dielectric-based acceleration or plasma wakefield-based acceleration (PWFA) may be available. 
For instance, in PWFA a region in a plasma is depleted of free electrons using a traversing particle drive beam, causing high electric fields to build up.
Acceleration gradients reaching up to 50 billion volts per metre (GV/m) -- 500 times larger than the CLIC gradient -- have already been demonstrated over a metre using this technique [26]. 
Although further studies are needed for these technologies to reach the beam quality and power efficiency needed for particle physics, advances in this area could allow realisation of much more compact linear accelerator designs. 

In the context of the CLIC project, these novel acceleration technologies are considered for application within and beyond the CLIC baseline programme [3]: 
an $\Pep\Pem$ accelerator reaching centre-of-mass energies of 10\,TeV and beyond would become conceivable. 
Operation at 10\,TeV allows for a significant improvement in the Higgs-boson self-coupling determination. 
At a few tens of TeV even triple-Higgs boson production becomes accessible. 
Reaching high luminosities in such a collider would impose unprecedented requirements on the beam-line alignment and stability, challenges for which CLIC operation will be an ideal testbed.

Special care is taken to ensure that the CLIC baseline infrastructure is compatible with the use of the novel technologies discussed above, for instance in terms of the beam crossing angle and the laser-straight accelerator layout [3]. 
A possible implementation at CLIC could include replacing parts or the full main linac with dielectric or plasma-based acceleration units. 
Furthermore, the CLIC drive-beam complex could be adapted for use in a beam-driven PWFA. 
The CLIC main linac injector complex, providing 9 GeV electrons and positrons, could be reused to inject directly into the new main linac.
\section{Outlook}
The CLIC project has an attractive timescale and a physics programme that is highly complementary to what can be learned at HL-LHC. 
With its first stage, CLIC provides a rich and guaranteed physics programme, after which the physics and technology landscape can be re-evaluated to allow for a physics motivated decision on the subsequent large-scale particle-physics facility. 
If the initial stage of CLIC at 380 GeV is endorsed in the 2020 update of the European Strategy for Particle Physics, the next step towards realisation of CLIC will be a preparation phase aiming to produce a Technical Design Report (TDR) by 2025.
\section*{References}
[1] ATLAS collaboration, Aad, G. et al., Observation of a new particle in the search for the Standard Model Higgs boson with the ATLAS detector at the LHC, Phys. Lett. B716, 1--29, (2012). 
\newline
[2] CMS collaboration, Chatrchyan, S. et al., Observation of a new boson at a mass of 125 GeV with the CMS experiment at the LHC, Phys. Lett. B716, 30--61, (2012).
\newline
[3] CLIC and CLICdp collaborations, Burrows, P.N. et al. (eds.), The Compact Linear Collider (CLIC) - 2018 Summary Report, CERN Yellow Reports: Monographs, 2, 1--98, (2018).
\newline
[4] CLIC accelerator collaboration, Aicheler, M. et al. (eds.), The Compact Linear Collider (CLIC) - Project Implementation Plan, CERN Yellow Reports: Monographs, 4, 1--247, (2018).
\newline
[5] de Blas, J. et al., The CLIC Potential for New Physics, CERN Yellow Reports: Monographs, 3, 1--274, (2018).
\newline
[6] Dannheim, D. et al. (eds.), Detector Technologies for CLIC, CERN Yellow Reports: Monographs, 1, 1--140, (2019).
\newline
[7] Candel, A. et al., Numerical Verification of the Power Transfer and Wakefield Coupling in the CLIC Two-Beam Accelerator, Conf.Proc. C110328, 51--55, (2011).
\newline
[8] CERN, Animation of the CLIC acceleration complex, \\\href{http://dx.doi.org/10.17181/cds.2688461}{\ttfamily http://dx.doi.org/10.17181/cds.2688461} (2019).
\newline
[9] CLIC and CLICdp collaborations, Boland, M.J. et al., Updated baseline for a staged Compact Linear Collider, CERN Yellow Reports, 4, 1--45, (2016).
\newline
[10] Apollinari, G. et al. (eds), High-Luminosity Large Hadron Collider (HL-LHC): Technical Design Report V. 0.1, CERN Yellow Reports: Monographs, 4, 1--516, (2017).
\newline
[11] Stapnes, S., The Compact Linear Collider, Nature Reviews Physics, 1, 235--237 (2019).
\newline
[12] CLICdp collaboration, Arominski, D. et al., A detector for CLIC: main parameters and performance, Preprint at \href{https://arxiv.org/abs/1812.07337} {\ttfamily https://arxiv.org/abs/1812.07337} (2018).
\newline
[13] Latina, A., Schulte, D., Stapnes, S., CLIC study update August 2019, CERN-ACC-2019-0051, Preprint at \href{http://cds.cern.ch/record/2687090}{\ttfamily http://cds.cern.ch/record/2687090} (2019).
\newline
[14] Linssen, L. et al. (eds.), Physics and Detectors at CLIC: CLIC Conceptual Design Report, CERN Yellow Reports: Monographs, CERN-2012-003, 1--257, (2012).
\newline
\newline
[15] Thomson, M.A., Particle Flow Calorimetry and the PandoraPFA Algorithm, Nucl.Instrum.Meth. A611 (2009) 25--40, (2009).
\newline
[16] CMS collaboration, The Phase-2 Upgrade of the CMS Endcap Calorimeter, CERN-LHCC-2017-023, Preprint at \href{http://cds.cern.ch/record/2293646}{\ttfamily http://cds.cern.ch/record/2293646} (2017).
\newline
[17] Sefkow, F. et al., Experimental Tests of Particle Flow Calorimetry, Rev.Mod.Phys. 88, 015003, (2016).
\newline
[18] FCAL collaboration, Abramowicz, H. et al., Performance and Moli\`{e}re radius measurements using a compact prototype of LumiCal in an electron test beam, Eur. Phys. J. C79 no.7, 579, (2019).
\newline
[19] Abramowicz, H. et al., Higgs physics at the CLIC electron-positron linear collider, Eur. Phys. J. C77 (2017) no.7, 475, (2017).
\newline
[20] CLICdp collaboration, Abramowicz, H., et al., Top-quark physics at the CLIC electron-positron linear collider, JHEP 11, 003, 1--84, (2019).
\newline
[21] de Blas, J. et al., Higgs Boson Studies at Future Particle Colliders, \\Preprint at \href{https://arxiv.org/abs/1905.03764}{\ttfamily https://arxiv.org/abs/1905.03764} (2019).
\newline
[22] Roloff, P. et al., Double Higgs boson production and Higgs self-coupling extraction at CLIC, Preprint at \href{https://arxiv.org/abs/1901.0589}{\ttfamily https://arxiv.org/abs/1901.0589} (2019).
\newline
[23] Cepeda, M. et al., Higgs Physics at the HL-LHC and HE-LHC, CERN-LPCC-2018-04, Preprint at \href{https://arxiv.org/abs/1902.00134}{\ttfamily https://arxiv.org/abs/1902.00134} (2019).
\newline
[24] Gupta, R. S., Rzehak, H. and Wells, J. D., How well do we need to measure Higgs boson couplings?, Phys. Rev., D86, 095001, (2012).
\newline
[25] European Strategy for Particle Physics Preparatory Group, Ellis, R.K. et al., Physics Briefing Book, CERN-ESU-004, Preprint at \href{ https://arxiv.org/abs/1910.11775}{\ttfamily https://arxiv.org/abs/1910.11775} (2019).
\newline
[26] Blumenfeld, I. et al., Energy doubling of 42 GeV electrons in a metre-scale plasma wakefield accelerator, Nature 445, 741--744, (2007). 
\section*{Acknowledgements}
Detailed studies of the CLIC accelerator design, physics potential, detector design, and R\&D on detector technologies are performed by the CLIC accelerator and CLIC detector and physics (CLICdp) collaborations. 
The CLIC project plans have recently been summarized in several comprehensive reports as input to the European Strategy for Particle Physics Update process\\ (\href{https://clic.cern/european-strategy}{\ttfamily https://clic.cern/european-strategy}). 
The authors gratefully acknowledge all contributors, authors and editors of these publications. 
In addition, the authors would like to thank Konrad Elsener, Lucie Linssen, Aidan Robson, Philipp Roloff, Ulrike Schnoor, Steinar Stapnes and Walter Wuensch who provided further essential input to this publication as well as Marko Petri\v{c} and Mateus Vicente Barreto Pinto for their contributions to the figures of this publication.

\end{document}